\documentclass{article}
\usepackage{spconf,amsmath,graphicx}
\usepackage{cuted} 
\usepackage{caption} 
\usepackage{subcaption} 
\usepackage{enumitem} 
\usepackage{float} 
\usepackage[colorlinks=true, linkcolor=blue, citecolor=blue, urlcolor=blue]{hyperref}
\hypersetup{pdfcreator  = {LaTeX with hyperref package},
               pdfproducer = {dvips + ps2pdf} }

\title{rate-distortion optimization with alternative references for \\UGC video compression}
\name{Xin Xiong$^*$, Eduardo Pavez$^*$, Antonio Ortega$^*$, Balu Adsumilli$^\dagger$ \thanks{This work was funded in part by a gift from YouTube}}
\address{University of Southern California, Los Angeles, CA USA \\ Google Inc, Mountain View, CA USA}
\begin{document}
\ninept
\maketitle
\begin{abstract}
User generated content (UGC) refers to videos that are uploaded by users and shared over the Internet. 
UGC may have low quality due to noise and previous compression. When re-encoding UGC for streaming or downloading, a traditional video coding pipeline will perform rate-distortion (RD) optimization to choose coding parameters. 
However, in the UGC video coding case, since the input is not pristine,  quality  ``saturation'' (or even degradation) can be observed, i.e., increased bitrate only leads to improved representation of coding artifacts and noise present in the UGC input.
%
In this paper, we  study the saturation problem in UGC compression, where the goal is to identify and avoid during encoding,  the coding parameters and rates that lead to quality saturation.
We proposed  a geometric  criterion for saturation detection  that works with  rate-distortion optimization, and  only requires a few frames from the  UGC video.  
In addition, we show how to combine the proposed saturation detection method with existing video coding systems that implement rate-distortion optimization for efficient  compression of UGC videos.
\end{abstract}
%



\section{Introduction}
\begin{figure*}[htb]
    \centering
    \begin{subfigure}[b]{0.24\textwidth}
    \includegraphics[width=\textwidth]{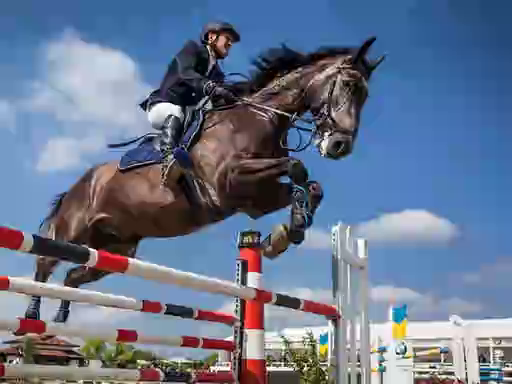}
    \caption{}
    \label{fig:S1_UGC}
    \end{subfigure}
    \hfill
    \begin{subfigure}[b]{0.24\textwidth}
    \includegraphics[width=\textwidth]{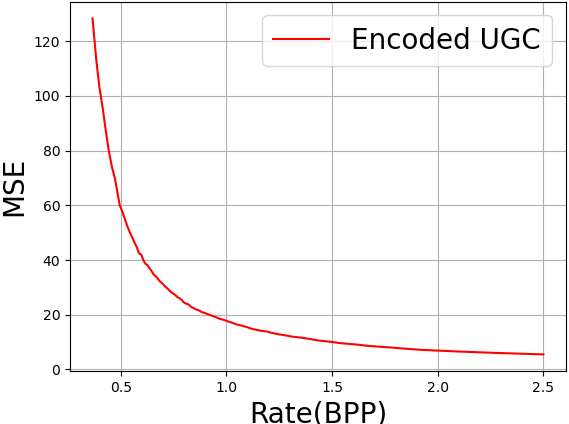}
    \caption{}
    \label{fig:S1_MSE}
    \end{subfigure}
    \hfill
    \begin{subfigure}[b]{0.24\textwidth}
    \includegraphics[width=\textwidth]{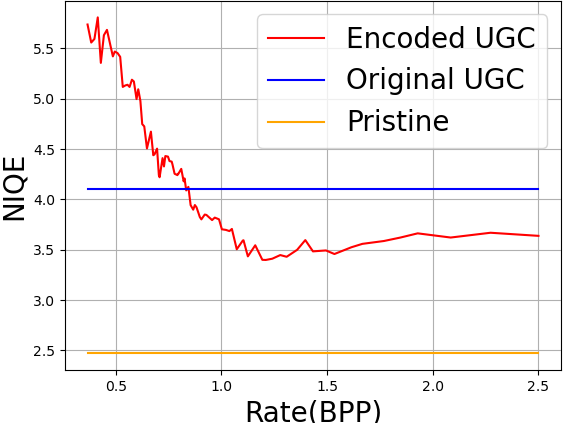}
    \caption{}
    \label{fig:S1_NIQE}
    \end{subfigure}
    \hfill
    \begin{subfigure}[b]{0.24\textwidth}
    \includegraphics[width=\textwidth]{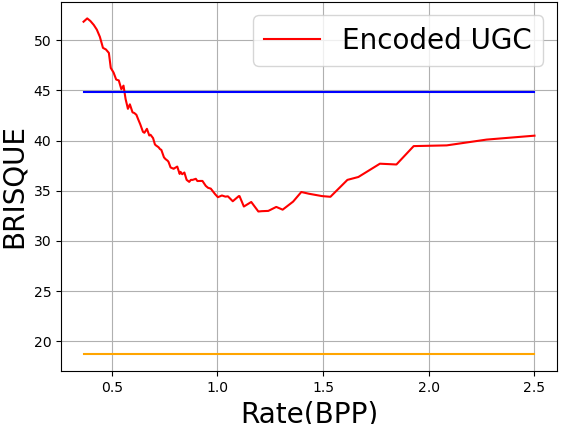}
    \caption{}
    \label{fig:S1_BRISQUE}
    \end{subfigure}
    \hfill
    \caption{
    An example of UGC compression problem. (a) A UGC image generated from KADID-10k\cite{kadid10k}. (b) Rate and distortion curve for encoding UGC. (c) and (d) rate and quality of encoded UGC. NIQE and BRISQUE are two popular no-reference quality metrics.}
    \label{fig:S1_intro}
\end{figure*}

Videos that are created by users and then uploaded to video sharing services (e.g. YouTube) are commonly referred to as user generated content (UGC) \cite{wang2019youtube}. Service providers \textit{re-encode} UGC videos at multiple resolutions and quality levels for streaming. 
The traditional video coding pipeline assumes the input video is pristine and makes  use of rate-distortion (RD) optimization to ensure efficient bitrate allocation \cite{ortega1998rate, sullivan1998rate}.
However, when encoding  UGC videos, which may  have low quality due to previous compression, a traditional RD optimization (RDO) framework that uses the UGC input as reference will allocate bitrate to preserve undesirable artifacts. 
To illustrate this idea consider the example of 
\autoref{fig:S1_intro} where a previously compressed image is re-encoded with JPEG and two no-reference image quality assessment metrics, NIQE \cite{mittal2012making} and BRISQUE \cite{mittal2012no}, are used to evaluate the perceptual quality of the re-encoded image. As shown by \autoref{fig:S1_NIQE} and \autoref{fig:S1_BRISQUE}, the perceptual  quality does not improve (i.e., it ``saturates'') when the rate is $R \approx 1.25  bpp$. 
While perceptual quality saturates,  mean-squared-error (MSE) continues to decrease (\autoref{fig:S1_MSE}), 
which suggests that additional rate is mostly helping represent coding artifacts in the UGC. Thus for this specific UGC image, rate should not be increased beyond $1.25 bpp$. 

Following \cite{pavez2022compression}, we define the UGC compression problem as: 
\emph{Given an UGC signal and a compression system (e.g., JPEG, H.264), choose coding parameters to accurately represent the perceptually meaningful parts of the signal, while avoiding allocating resources to encode compression artifacts and noise.}    
Recent approaches to choose coding parameters for UGC are based on no-reference quality metrics  and content quality  classification approaches \cite{mittal2012no, wang2021rich,john2020rate, ling2020towards,tu2021efficient,yu2021predicting,wang2020video}. 
However, these strategies are out-of-the-loop (i.e., they are applied before encoding or after decoding, as in the example from \autoref{fig:S1_intro}), and usually require training on a large-scale dataset. 
In \cite{pavez2022compression}, we addressed the UGC compression problem from an RD theoretic perspective, where the signal to be encoded is noisy and  distortion is computed  with respect to an unavailable pristine reference. Using the noisy source coding theorem \cite{dobru1962source,berger1971rate}, we showed that a theoretically optimal UGC compression system  applies an optimal denoiser to the UGC video followed by optimal source coding of the denoised UGC. 
For a practical UGC video compression system,  
we proposed using off-the-shelf denoisers to compute a denoised UGC to be used as an alternative reference for distortion computation. We  showed that quality saturation  occurs, and proposed a method to detect it. 
In this paper, we address several of the limitations of our previous approach \cite{pavez2022compression} and work towards  developing a practical UGC compression system.
While in theory  
a denoised UGC should be used both as a reference and as a source, in \cite{pavez2022compression} it was only used as a reference. In this work, we provide further evidence to support that choice. We show  that there is little perceptual difference between encoding the UGC and encoding the denoised UGC (see \autoref{ssec:exp_denoiseUGCenco}). In addition, the RD curve from encoding denoised UGC does not have a MSE saturation region, which makes it difficult to decide when to stop increasing the bitrate. Finally, encoding a denoised UGC requires denoising the whole UGC video which comes with a high computational cost, while the proposed UGC compression approach only requires denoising a few video frames for saturation detection.
%
We also show that the saturation criterion in \cite{pavez2022compression} does not always work and depends on the denoising method. As an alternative, we propose a new saturation detection criterion (\autoref{ssec:method_saturation}). Our approach is more principled, while also being more robust to the choice of denoiser. 
%
Finally, we propose a low complexity saturation detection method for the whole video by performing RDO with a simplified codec, and using only a subset of frames (\autoref{ssec:method_rdo}).  Our method returns a \emph{saturation parameter} ($\lambda^*$) that controls the RD trade-off and can be given as input to a traditional video coding system that performs the actual encoding and RDO (see \autoref{fig:Intro_proposed_system}). Our proposed method serves as a practical plug-in module that can operate with any codec implementing RDO.

\section{Quality saturation and RDO }
In this section we  describe our pipeline. We first introduce an elementary inequality that relates various distortion metrics and use it to  define a saturation region. Then we describe the proposed UGC compression system that performs RDO. 
\subsection{Geometric relation between different distortions}
\label{ssec:method_geometrydistortions}
We consider a traditional video compression system that encodes the UGC video $\mathbf{U}$ with parameters $\theta$, and produces a decoded UGC video  $\hat{\mathbf{U}}(\theta)$. For such a system, there exist a set of parameters $\Theta=\lbrace \theta_1, \theta_2,\cdots \rbrace$, so that $\lim_{k \rightarrow \infty}\Vert \mathbf{U} - \hat{\mathbf{U}}(\theta_k)\Vert = 0$, and thus $\hat{\mathbf{U}}(\theta_k) \rightarrow \mathbf{U}$, and the rate of the encoded representation of $\hat{\mathbf{U}}(\theta_k)$ increases with $k$. Examples of  $\theta_k$ include QP parameters used in H.264.
A traditional video coding system, would use the distortion $\Vert \hat{\mathbf{U}}(\theta_k) - \mathbf{U} \Vert^2$ to measure quality and perform RDO. In \cite{pavez2022compression} we proposed using $ \Vert \hat{\mathbf{U}}(\theta_k) - \mathbf{Z} \Vert^2$ to compute distortion, where $\mathbf{Z}$ is a de-noised signal obtained from $\mathbf{U}$.  We showed experimentally that the corresponding RD curve saturates at higher bitrates, and  $ \Vert \hat{\mathbf{U}}(\theta_k) - \mathbf{Z} \Vert^2$  converges to $\Vert \mathbf{U} - \mathbf{Z} \Vert^2$. 
%
%
 We can formalize this observation as follows. Using the identity $  \Vert \mathbf{x} - \mathbf{z} \Vert^2 - \Vert \mathbf{y} - \mathbf{z} \Vert^2  =   \Vert \mathbf{x} - \mathbf{y} \Vert^2 + 2\langle\mathbf{x} - \mathbf{y}  , \mathbf{y} - \mathbf{z}\rangle   $,  valid for any $\mathbf{x},\mathbf{y}, \mathbf{z}$,  and the triangular inequality,  the  distortions  are related as follows:
\begin{align}\label{eq_mse_ineq1}
   &\left\vert  \Vert \hat{\mathbf{U}}(\theta) - \mathbf{Z} \Vert^2 - \Vert \mathbf{U} - \mathbf{Z} \Vert^2 \right\vert \leq   \\ \label{eq_mse_ineq2}
   & \Vert \hat{\mathbf{U}}(\theta) - \mathbf{U} \Vert^2  +2\Vert \hat{\mathbf{U}}(\theta) - \mathbf{U} \Vert\Vert \mathbf{U}- \mathbf{Z} \Vert
\end{align}
for any reference $\mathbf{Z}$ and any coding parameter $\theta \in \Theta$. 
From \eqref{eq_mse_ineq2}, we can see that convergence occurs faster at low rates than at high rates, and detecting the transition from one regime to the other will be helpful to detect saturation. More precisely, we observe:

\noindent\textbf{High bitrate, low distortion regime.}    When $\Vert \hat{\mathbf{U}}(\theta) - \mathbf{U} \Vert$ is small,  then   $\vert  \Vert \hat{\mathbf{U}}(\theta) - \mathbf{Z} \Vert^2 - \Vert \mathbf{U} - \mathbf{Z} \Vert^2 \vert= \mathcal{O}(\Vert \hat{\mathbf{U}}(\theta) - \mathbf{U} \Vert) $. In this region, distortion decays more slowly and thus quality has saturated.
    
\noindent\textbf{Low bitrate, high distortion regime.}  When  $\Vert \hat{\mathbf{U}}(\theta) - \mathbf{U} \Vert$ is larger, then  $\vert  \Vert \hat{\mathbf{U}}(\theta) - \mathbf{Z} \Vert^2 - \Vert \mathbf{U} - \mathbf{Z} \Vert^2 \vert= \mathcal{O}(\Vert \hat{\mathbf{U}}(\theta) - \mathbf{U} \Vert^2) $. In this region, distortion decays at the same rate as a traditional system.
    
\noindent\textbf{Transition point.} Transition from quadratic decay (low bitrates) to linear decay (higher bitrates) occurs approximately  when both terms on the right side of \eqref{eq_mse_ineq2} are equal, i.e.,  $\Vert \hat{\mathbf{U}}(\theta) - \mathbf{U}  \Vert = 2\Vert \mathbf{U}  - \mathbf{Z}  \Vert$.
\begin{figure}[ht]
    \centering
    \includegraphics[width = 0.45\textwidth]{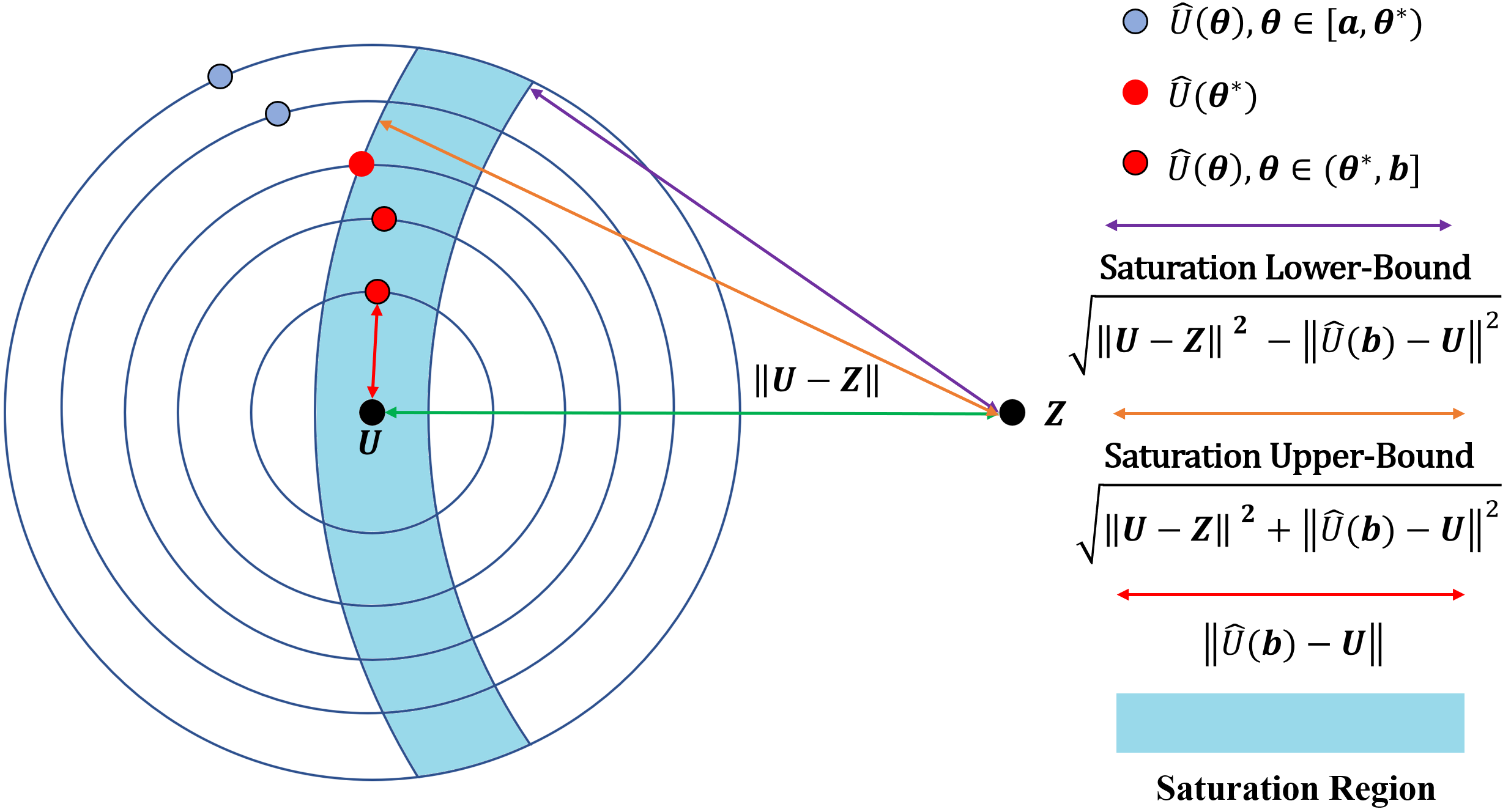}
    \caption{An illustration for our proposed saturation detection method. Each node represents a position of $\hat{\mathbf{U}}(\theta)$.
    The length of red line is the minimal distortion which can be achieved by $\theta = b$. 
    }
    \label{fig:Exp_Saturation}
\end{figure}
\subsection{Saturation region detection}
\label{ssec:method_saturation}
We now use the results from the previous section to detect saturation, i.e.,  when  $\hat{\mathbf{U}}(\theta)$ is sufficiently close to $\mathbf{U}$. 
From \eqref{eq_mse_ineq1} we see that in the saturation region,  distortion decreases more slowly. Thus  our goal will be to identify the transition point from quadratic to linear behavior. Without loss of generality, suppose the coding parameter $\theta$ belongs to a closed interval $\Theta=[a, b]$, with a higher $\theta$ resulting in higher bitrate and lower distortion. 
We define the saturation region $[\theta^*, b]$ as the largest interval (the one with the smallest $\theta^*$) such that:
\begin{equation}\label{theta_saturation}
\left\vert  \Vert \hat{\mathbf{U}}(\theta) - \mathbf{Z} \Vert^2 - \Vert \mathbf{U} - \mathbf{Z} \Vert^2 \right \vert \leq \Vert \hat{\mathbf{U}}(b) - \mathbf{U} \Vert^2, \forall \theta \in [\theta^*, b].
\end{equation}
We say that  $\theta^*$ is the saturation onset or saturation point. Note that 
\begin{align}
    \Vert \hat{\mathbf{U}}(b) - \mathbf{U} \Vert^2 &\leq  \Vert \hat{\mathbf{U}}(b) - \mathbf{U} \Vert^2  +2\Vert \hat{\mathbf{U}}(b) - \mathbf{U} \Vert\Vert \mathbf{U}- \mathbf{Z} \Vert \\ \label{theta_equa}
   &\leq  \Vert \hat{\mathbf{U}}(\theta) - \mathbf{U} \Vert^2  +2\Vert \hat{\mathbf{U}}(\theta) - \mathbf{U} \Vert\Vert \mathbf{U}- \mathbf{Z} \Vert
\end{align}
$\forall \theta \in [a, b]$,
thus \eqref{theta_saturation} is stronger than \eqref{eq_mse_ineq2}.
The region \eqref{theta_saturation} is illustrated in   \autoref{fig:Exp_Saturation}. Each circle centered at  $\mathbf{U}$ has radius $\Vert \hat{\mathbf{U}}(\theta) - \mathbf{U} \Vert$, representing possible positions of $\hat{\mathbf{U}}(\theta)$. 
The   radius of the smallest circle is $\Vert \hat{\mathbf{U}}(b) - \mathbf{U} \Vert$. 
The saturation region is delimited by two circles centered at $\mathbf{Z}$. 
The largest and smallest circles have equation 
\begin{align}
    \Vert \hat{\mathbf{U}}(\theta) - \mathbf{Z} \Vert^2 &= {\Vert \mathbf{U} - \mathbf{Z} \Vert^2 + \Vert \hat{\mathbf{U}}(b) - \mathbf{U} \Vert^2} = \Delta^2, \textnormal{ and } \\
    \Vert \hat{\mathbf{U}}(\theta) - \mathbf{Z} \Vert^2 &= {\Vert \mathbf{U} - \mathbf{Z} \Vert^2 - \Vert \hat{\mathbf{U}}(b) - \mathbf{U} \Vert^2} = \delta^2,
\end{align}
with radius $\Delta$ and $\delta$, respectively.
The blue and red points in \autoref{fig:Exp_Saturation} show how the saturation $\theta^*$ is detected, as we enter the shaded blue region.  As $\theta$ increases, $\hat{\mathbf{U}}(\theta)$ approaches $\mathbf{U}$, so that $\Vert \hat{\mathbf{U}}(\theta) -  \mathbf{U} \Vert$ is getting smaller and $\Vert \hat{\mathbf{U}}(\theta) -  \mathbf{Z} \Vert$ is saturating. 
%
\subsection{Rate-distortion optimization with alternative reference}
\label{ssec:method_rdo}
\begin{figure}[h]
    \centering
    \begin{subfigure}[b]{0.4\textwidth}
    \centering
    \includegraphics[width=\textwidth]{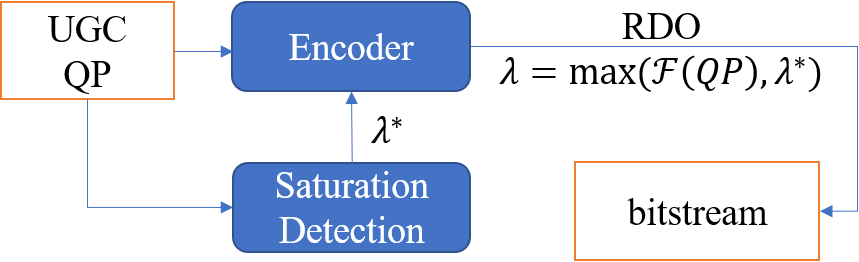}
    \caption{UGC compression system}
    \label{fig:S1_UGC}
    \end{subfigure}
    \vfill
    \begin{subfigure}[b]{0.4\textwidth}
    \centering
    \includegraphics[width=\textwidth]{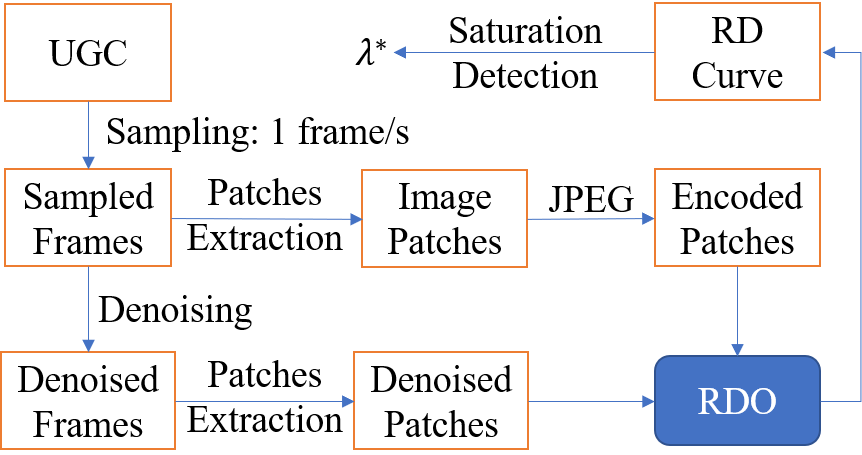}
    \caption{Saturation detection system}
    \label{fig:S2_UGC}
    \end{subfigure}
    \caption{
    An illustration of how our proposed method work with existing video codecs.
    (b) shows how our saturation detection method works with encoder as a practical plug-in module.
    (c) shows the details of our saturation detection method.
    }
    \label{fig:Intro_proposed_system}
\end{figure}

The encoder-decoder parameters in a traditional video codec are obtained using RDO \cite{ortega1998rate,sullivan1998rate} by solving the optimization problem 
\begin{equation} \label{eq:J}
    \min_{\boldsymbol\alpha}D(\boldsymbol\alpha) + \lambda R(\boldsymbol\alpha),
\end{equation}
where $J=D(\boldsymbol\alpha) + \lambda R(\boldsymbol\alpha)$ is the Lagrange cost, $\boldsymbol\alpha$ is a set of coding parameters (e.g., mode decision, quantization parameters, transform size, etc.), $D(\boldsymbol\alpha)$ is the MSE distortion between the encoded UGC $\hat{\mathbf{U}}(\boldsymbol\alpha)$ and the input $\mathbf{U}$. $R(\boldsymbol\alpha)$ is the rate of  $\hat{\mathbf{U}}(\boldsymbol\alpha)$. The parameter $\lambda>0$ controls the trade-off between rate and distortion. When $\lambda$ is small, solutions with higher bitrate are encouraged, and thus when encoding UGC, a small $\lambda$ leads to quality saturation. 
\autoref{fig:S1_UGC} depicts the proposed  UGC compression system. It consists of a traditional video compression system that performs RDO, along with a saturation detection module, which takes an input UGC video, and returns a saturation $\lambda^*$. In  a traditional compression system, a typical input parameter is the ``QP'' value which can be easily converted into an input $\lambda$ \cite{richardson2011h}. If $\lambda<\lambda^*$, the input ``QP'' parameter is in the saturation region, thus the encoder performs RDO with parameter $\lambda^*$. When $\lambda\geq\lambda^*$, the encoder performs RDO as normal with parameter $\lambda$. 
The saturation detection module depicted in \autoref{fig:S2_UGC} also utilizes RDO. Because saturation detection requires a RD curve covering a wide range of RD pairs, in order to reduce computational complexity we consider a simplified system where only a few frames are considered (e.g., $1$ per second). For encoding, we use  JPEG and allow different quantization parameters for each patch. 

The UGC video $\mathbf{U}$, and the denoised UGC $\mathbf{Z}$ are partitioned into $N$ non overlaping patches. For each patch,  there is a single coding parameter $\alpha_k$ that will be optimized. The rate  and distortion of the $k$th patch are denoted by $R(\alpha_k)$ and $D(\alpha_k)$, respectively.  $D(\alpha_k) = \Vert \hat{\mathbf{U}}_k(\alpha_k) -  \mathbf{Z}_k \Vert^2$ is the distortion between the encoded $k$th patch  $\hat{\mathbf{U}}_k(\alpha_k)$ and the denoised $k$th patch $\mathbf{Z}_k$. Given $\lambda>0$, we solve the RDO problem
\begin{equation}\label{eq_rdo_saturation}
    \min_{\alpha_1, \cdots, \alpha_N} \sum_{k=1}^N D(\alpha_k) + \lambda\sum_{k=1}^N R(\alpha_k).
\end{equation}
The vector of optimal coding parameters that solves \eqref{eq_rdo_saturation} is denoted by $\boldsymbol\alpha_{\lambda}$, while the optimal total rate is denoted by $R(\boldsymbol\alpha_{\lambda})$ and the optimal total distortion (MSE between encoded UGC and denoised UGC) is given by $D(\boldsymbol\alpha_{\lambda})$.
We apply RDO and solve \eqref{eq_rdo_saturation} for a wide range of $\lambda$ to obtain an RD curve with pairs $(D(\boldsymbol\alpha_{\lambda}), R(\boldsymbol\alpha_{\lambda}))$, from which saturation can be detected.
%
%
Given $\lambda \in [\lambda_{min}, \lambda_{max}]$ and $D(\boldsymbol\alpha_{\lambda})=\Vert \hat{\mathbf{U}}(\boldsymbol\alpha_{\lambda}) - \mathbf{Z} \Vert^2$, we can rewrite the saturation equation \eqref{theta_saturation} and define saturation $\lambda^*_Z$  as the largest value satisfying
\begin{equation}\label{lambdaZ_saturation}
\left\vert  \Vert \hat{\mathbf{U}}(\boldsymbol\alpha_{\lambda}) - \mathbf{Z} \Vert^2 - \Vert \mathbf{U} - \mathbf{Z} \Vert^2 \right \vert \leq \Vert \hat{\mathbf{U}}(\boldsymbol\alpha_{\lambda_{min}}) - \mathbf{U} \Vert^2, 
\end{equation}
for all $\lambda \in [\lambda_{min}, \lambda^*_Z]$.
In a traditional codec, distortion is computed with respect to $\mathbf{U}$. Since $\lambda^*_Z$ is the saturation we found when $\mathbf{Z}$ is the reference, it controls the rate-distortion trade-off using this particular choice of distortion.  We still need to find the saturation $\lambda^*_U$ appropriate for the case where $\mathbf{U}$ is the reference used for distortion computation, and that can be provided as input to a traditional codec performing RDO.  Once we obtain $\lambda^*_Z$,   the saturation rate is given by $ R(\boldsymbol\alpha_{\lambda^*_Z})$. As we do not want to operate at a rate larger than $ R(\boldsymbol\alpha_{\lambda^*_Z})$, we define the saturation $\lambda^*_U$   as the smallest value such that 
\begin{equation}\label{lambdaU_saturation}
R(\boldsymbol\beta_{\lambda}) \leq R(\boldsymbol\alpha_{\lambda^*_Z}), \quad  \lambda \in [\lambda^*_U,\lambda_{max}],
\end{equation}
where $\boldsymbol\beta_{\lambda}$ is the minimizer of the RDO of \eqref{eq_rdo_saturation} with parameter $\lambda$ and using the traditional distortion $D(\alpha_k) = \Vert \hat{\mathbf{U}}_k(\alpha_k) -  \mathbf{U}_k \Vert^2$.
\section{Experiment}
\subsection{Encoding the de-noised references}
\label{ssec:exp_denoiseUGCenco}
\begin{figure}[h]
    \flushleft
    \includegraphics[width=0.48\textwidth]{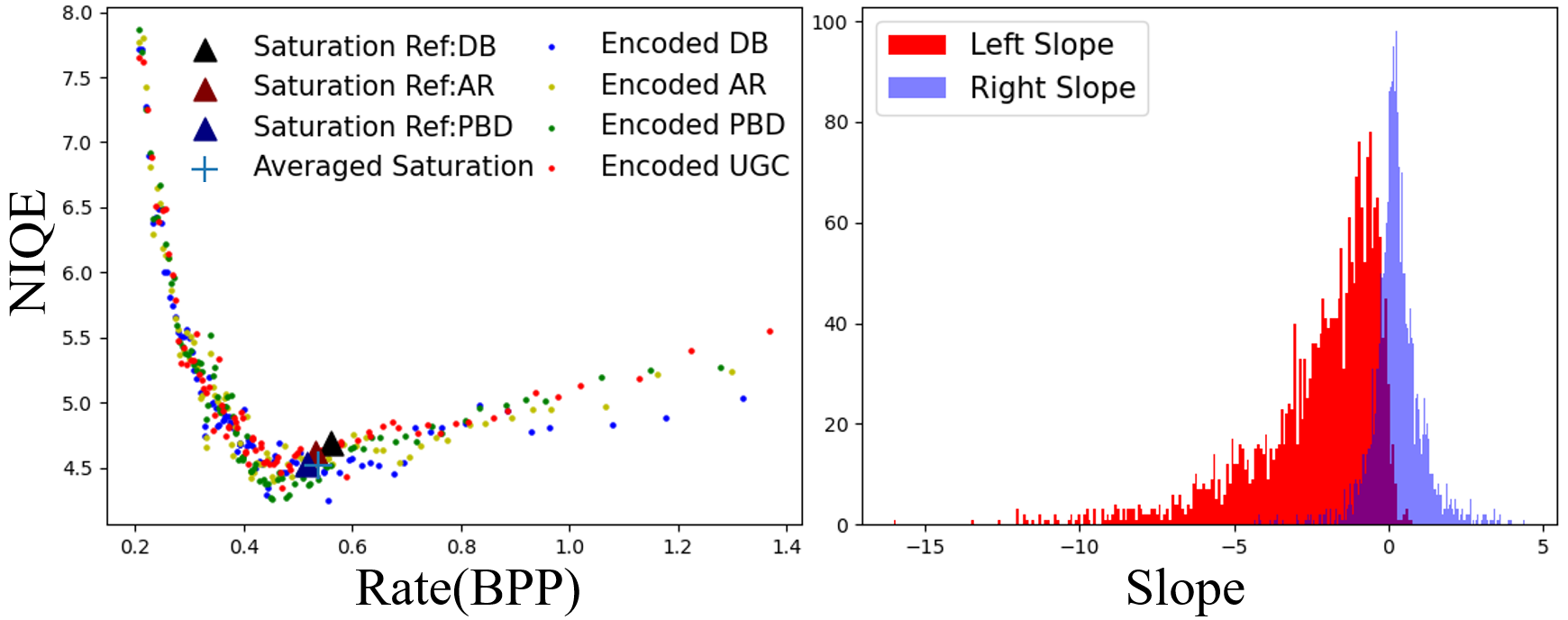}
    \caption{Left: Rate and NIQE curves for encoding UGC image ('I80', $QP=45$) and de-nosied UGC at different JPEG $QV$. 
    Right: Histogram of slopes. 'Left/Right slope' is average slope between average saturation point and points left/right to it.  
    Each UGC has both 'left' and 'right' slope. 
    The histogram is based on $2349$ images.
    }
    \label{fig:Histogram}
\end{figure}

For this experiment, we use the KADID-10k dataset \cite{kadid10k, deepfl-iqa}, which has $81$ pristine images named from 'I01' to 'I81'. For each pristine image, we used the H.264 in Ffmpeg to simulate UGC images with compression noise  using quantization values from $23$ to $51$. 
Then, for each UGC image, we applied three different video/image restoration methods: De-Blocking (DB) from ffmpeg, Artifact Removal \cite{jiang2021towards} (AR) and Practical Blind Denoising \cite{zhang2022practical} (PBD). 
We encode each image (UGC or de-noised UGC) with JPEG \cite{clark2015pillow}, using $86$ different quality values ($QV = 10, 11, ..., 95$). 
Then, we use Naturalness Image Quality Evaluator (NIQE) \cite{mittal2012making} to evaluate the quality of the compression results. 
Note that we could still apply the idea from \autoref{fig:Exp_Saturation} to detect the saturation. Since RDO is not used in this experiment, we detect the saturation $\theta^*$ (JPEG $QV^*$) instead of $\lambda$ using constraint \eqref{theta_saturation}. 
Choosing different $\mathbf{Z}$, we may obtain different saturation points as is shown in  \autoref{fig:Histogram} left.
Also, we mark the location of averaged saturation point considering all the saturation points detected by different $\mathbf{Z}$ with crossing. 
There are very clear turning points on the curves, which indicates that the saturation behavior exists when encoding either UGC or de-noised references. 
To further understand the saturation phenomenon, we calculated the average slope between the detected saturation point and  points to its left and its right. 
As is shown in \autoref{fig:Histogram} right, the left slopes are almost all negative (since quality improves before saturation) while the right slopes appear randomly distributed around zero. This provides evidence  of successful saturation detection.

\begin{figure}[h]
    \centering
    \includegraphics[width=0.48\textwidth]{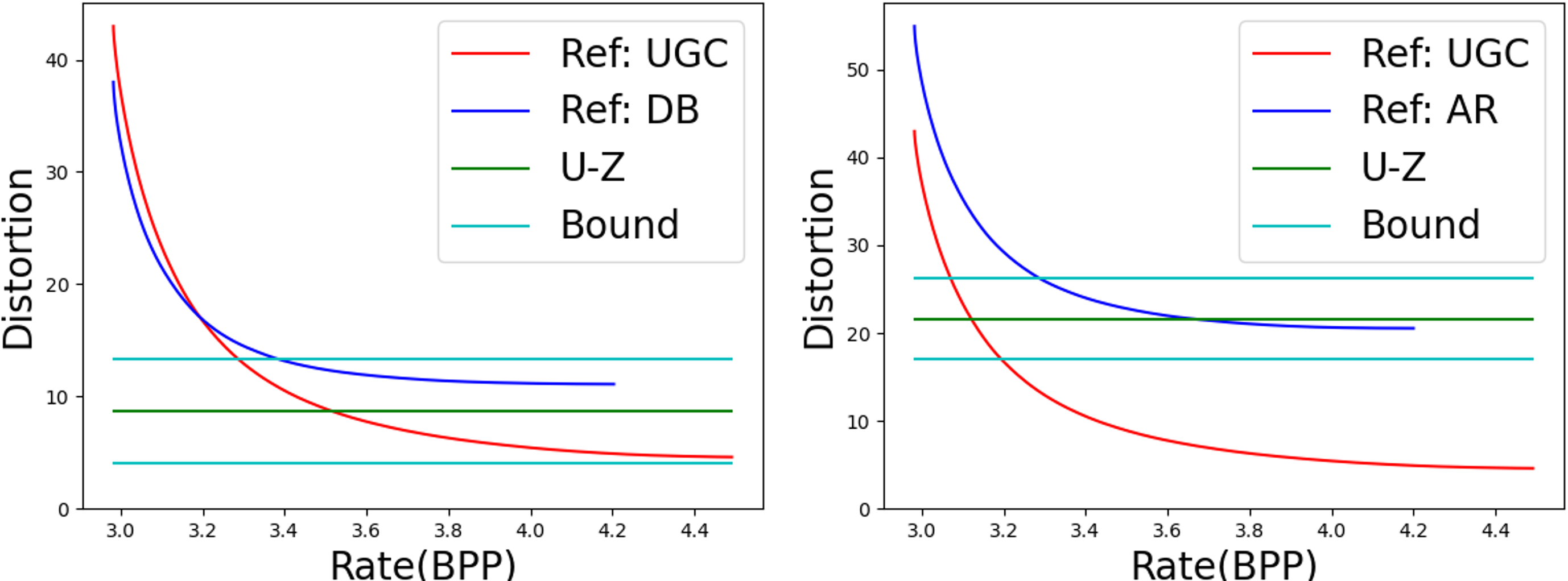}
    \caption{RD curves for `S11' UGC video sequence with $QP=35$. In Denoised references are DB (left) and AR (right). In each figure, the red RD curve uses $\mathbf{U}$ as  reference, while the blue RD curve uses $\mathbf{Z}$ as  reference. The green line represents the distortion between $\mathbf{U}$ and $\mathbf{Z}$. The two light blue lines are the upper and lower bounds from the proposed saturation detection method \eqref{lambdaZ_saturation}. }
    \label{fig:RD_Curve_}
\end{figure} 
\begin{figure}[h]
   \centering
    \includegraphics[width=0.48\textwidth]{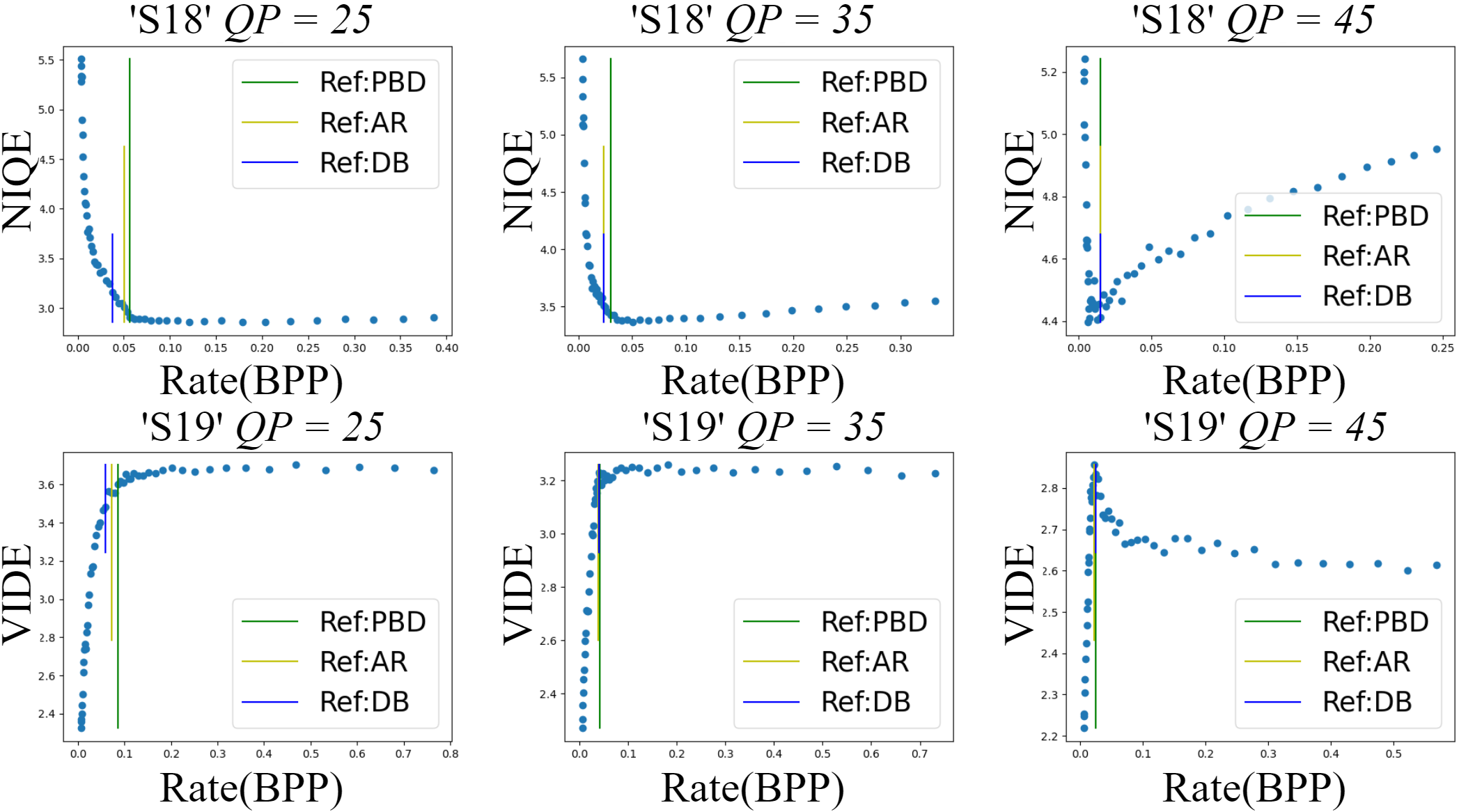}
    \caption{Video quality evaluation versus rate. We show  results for 6 UGC videos generated from 'S18' and 'S19' with 3 different $QP$. In each figure, blue points represent encoded UGC with H.264 at different rates. 
    The vertical lines represents the rate when re-compressing the UGC video with detected $QP(\lambda^*_U)$ using different $\mathbf{Z}$. 
    }
    \label{fig:VQA}
\end{figure}



\subsection{Saturation Detection on UGC Videos}

The video dataset BVI-CC \cite{katsenou2019subjective, zhang2020comparing} contains 9 pristine video sequences (named 'S11' to 'S19') with two resolutions: $3840 \times 2160$ and $1920 \times 1080$. Each video is $5$ seconds long, with $60$ frames per second.
For convenient, we used the $1920 \times 1080$ videos and down-sampled them into $480 \times 360$.
We took the down-sampled videos as the pristine videos and used  H.264 in Ffmpeg to create UGC videos with $QP$ ranging from 23 to 51. 
Then we uniformly sampled 5 frames out of 300 for saturation detection. 
On each video, we applied three different image/video restoration methods: De-Blocking (DB) from Ffmpeg, Artifacts Removal \cite{jiang2021towards} (AR) and Practical Blind Denoising \cite{zhang2022practical} (PBD). 
Then, each frame and its DB, AR and PBD references were uniformly divided into 90 patches of size $48 \times 40$. 
We encoded each patch with JPEG \cite{clark2015pillow}, using   $20$ different quality values ($QV_n$, $n=1,\cdots, 20$) ranging from from $19$ (worst) to $95$ (best).
Since JPEG has only one quality factor, for each patch, the minimization of RDO object function \eqref{eq:J} could be easily achieved by traversing all possible $QV_n$. 
\autoref{fig:RD_Curve_} shows the RD curves of a UGC video ($QP=35$) with two different de-noised references, DB and AR. In each figure, the red curve takes $\mathbf{U}$ as reference while the blue curve takes $\mathbf{Z}$ as reference.
Each point on a curve represents compressing patches by JPEG via RDO toward a specific $\lambda$. 
\autoref{fig:RD_Curve_} shows that blue curves converge to the green line, which is $||\mathbf{U} - \mathbf{Z} ||^2$. Using the proposed saturation detection method \eqref{lambdaZ_saturation}, the saturation $\lambda^*_Z$ could be found at the intersection of the blue curve and the upper light blue line, which is the upper bound of saturation in \eqref{lambdaZ_saturation}. Then, we could find the $\lambda^*_U$ from the red curve using \eqref{lambdaU_saturation}. 
In \cite{pavez2022compression}, saturation is detected by the intersection of blue and red RD curves.
However, when $\mathbf{Z}$ is computed with AR, the blue curves are always under the red curves and there is no intersection, thus the saturation detection method from\cite{pavez2022compression} would fail in this case. 
Nevertheless, the saturation exists and could be detected by our method. 
\subsection{Experiments with real H.264 codec}
To evaluate the saturation detected, we re-compressed the UGC video with Ffmpeg H.264 codec using $52$ different $QP$ and evaluated the results. 
Since Ffmpeg does not accept $\lambda$ as an input (which is the output of our saturation detection method),  we use
 the following relation \cite{richardson2011h}, valid for  $QP = 0, 1, ..., 51$,
 \begin{equation}
   \lambda(QP) = 0.852 \times 2^{(QP - 12)/3},
 \end{equation}
to translate the saturation $\lambda^*_U$ in to $QP(\lambda^*_U)$. 
Choosing a $QP < QP(\lambda^*_U)$  will result in waste of rate and no improvement on quality. 
We use NIQE and a no-reference quality metric for UGC video, VIDE \cite{tu2021ugc} for quality evaluation. 
Note lower NIQE value means better quality while higher VIDE value represents better quality. \autoref{fig:VQA} shows the quality of re-compressed UGC videos H.264 at different rates.
For any UGC video in \autoref{fig:VQA}, we could see clearly the saturation happens with both of the NIQE and VIDE. 
Furthermore, our detected saturation $\lambda^*_U$ accurately marks the turning point of the quality curves, which shows the correctness and accuracy of our proposed method.
We also see that when $QP = 25$ and $\mathbf{Z}$ is DB, the saturation detected (blue line) is worse than $\mathbf{Z}$ is AR or PBD. This indicates that DB may not be adaptive enough to UGC input with good quality, while AR and PBD produce better results.

\section{Conclusion}
In this paper, we proposed a practical system for quality saturation detection that can be combined with traditional compression systems for  UGC  video compression. 
The  proposed   saturation detection is based on   RDO  implemented on a simplified low complexity encoder. We illustrate the effectiveness of the proposed method using H.264 to compress previously compressed UGC. Our results show that the quality saturation region detected by our method is consistent with the saturation region found using no reference video quality metrics. 
We also show that our saturation detection method is robust to the choice of denoiser. However, our saturation detection fails when the input UGC quality is very low, and thus there is no hope that a good denoiser exists. When the UGC quality is high, and thus no saturation is expected,  the proposed method fails when the denoising algorithm does not set its parameters correctly and over smooths the input UGC. The proposed  framework can be complement and enhanced by exploiting  existing  approaches for quality detection, content type classification and no reference visual quality metrics. 

\clearpage

\bibliographystyle{IEEEbib}
\bibliography{refs}

\end{document}